\documentclass{article}

\usepackage{PRIMEarxiv}

\usepackage[utf8]{inputenc} % allow utf-8 input
\usepackage[T1]{fontenc}    % use 8-bit T1 fonts
\usepackage{hyperref}       % hyperlinks
\usepackage{url}            % simple URL typesetting
\usepackage{booktabs}       % professional-quality tables
\usepackage{amsfonts}       % blackboard math symbols
\usepackage{nicefrac}       % compact symbols for 1/2, etc.
\usepackage{microtype}      % microtypography
\usepackage{fancyhdr}       % header
\usepackage{graphicx}       % graphics
\usepackage{subfig}
\usepackage{multirow}
\usepackage{enumerate}
\usepackage{amsmath,amssymb}
\usepackage{tikz}

\graphicspath{{media/}}     % organize your images and other figures under media/ folder

\newcommand{\mycircle}[1]{\tikz{\filldraw[draw=#1,fill=#1] (0,0) circle [radius=0.1cm];}}

\newcommand{\mytriangle}[1]{\tikz{\filldraw[draw=#1,fill=#1] (0,0) -- (0.2cm,0) -- (0.1cm,0.2cm);}}

\newcommand{\mysquare}[1]{\tikz{\filldraw[draw=#1,fill=#1] (0,0) rectangle (0.2cm,0.2cm);}}

\definecolor{mygreen}{RGB}{34,139,34}
\definecolor{myred}{RGB}{178,34,34}

\title{RFID-based Article-to-Fixture Predictions in Real-World Fashion Stores}

\author{
  Matthias W{\"o}lbitsch \\
  Detego GmbH \\
  Graz, Austria \\
  \texttt{m.woelbitsch@detego.com} \\
  \And
  Thomas Hasler \\
  Detego GmbH \\
  Graz, Austria \\
  \texttt{t.hasler@detego.com} \\
  \And
  Patrick Kasper \\
  Detego GmbH \\
  Graz, Austria \\
  \texttt{p.kasper@detego.com} \\
  \And
  Denis Helic \\
  Graz University of Technology \\
  Graz, Austria \\
  \texttt{dhelic@tugraz.at} \\
  \And 
  Simon Walk \\
  Detego GmbH \\
  Graz, Austria \\
  \texttt{s.walk@detego.com} \\
}

\begin{document}
\maketitle

\begin{abstract}
In recent years, Radio Frequency Identification (RFID) technology has been applied to improve numerous processes, such as inventory management in retail stores.
However, automatic localization of RFID-tagged goods in stores is still a challenging problem. 
To address this issue, we equip fixtures (e.g., shelves) with reference tags and use data we collect during RFID-based stocktakes to map articles to fixtures. 
Knowing the location of goods enables the implementation of several practical applications, such as automated Money Mapping (i.e., a heat map of sales across fixtures).
Specifically, we conduct controlled lab experiments and a case-study in two fashion retail stores to evaluate our article-to-fixture prediction approaches.
The approaches are based on calculating distances between read event time series using DTW, and clustering of read events using DBSCAN.
We find that, read events collected during RFID-based stocktakes can be used to assign articles to fixtures with an accuracy of more than 90\%.
Additionally, we conduct a pilot to investigate the challenges related to the integration of such a localization system in the day-to-day business of retail stores.
Hence, in this paper we present an exploratory venture into novel and practical RFID-based applications in fashion retails stores, beyond the scope of stock management.
\end{abstract}


\section{Introduction}

Radio Frequency Identification (RFID) technology is widely adopted in the fashion retail industry for inventory management along the supply chain~\cite{moon2008, garrido2012, cilloni2019}.
RFID tagging of fashion goods with passive tags during the production enables retailers to keep track of individual items during distribution, as well as inside their brick-and-mortar stores.
A typical RFID setup in fashion retail stores consists of a handheld reader allowing store staff to perform daily RFID-enabled stocktakes.
This drastically improves inventory accuracy~\cite{hardgrave2009}, as stocktakes are performed quickly in short and regular intervals. 

\paragraph{Problem}
However, although RFID technology is already used by many retailers to improve their business processes, it is rarely used for the localization of goods inside stores.
While there exist several RFID-based localization approaches~\cite{ni2003, alippi2006, grisetti2007, xiao2017}, most of them require additional hardware, infrastructure, or parameter tuning, which is not feasible for most retailers due to the associated efforts, costs, and required expertise.
Nevertheless, knowing the location of goods is beneficial for several real-world applications.
For example, linking sales figures of articles to their placement within a store (i.e., \textit{Money Mapping}) allows visual merchandising departments to better understand the behavior of their customers~\cite{rizzi2017}.
In fact, many store owners today evaluate their visual merchandising strategies by manually performing Money Mapping.
However, this is a tedious and time consuming task---especially long-term---due to the constant changes that take place in a fashion store, and is usually only executed for a subset of fixtures (e.g., for individual shelves or tables) within a limited time frame.

\paragraph{Approach}
Therefore, we set out to tackle the problem of automatically determining locations of articles in fashion retail stores while keeping the administrative and financial overhead at a minimum.
To that end, we use read events during commonly performed RFID-based stocktakes, in combination with strategically placed passive RFID reference tags, to infer locations of articles on fixture-level, based on temporal clustering and time series similarities.
Hence, we follow a data-driven approach that requires no additional hardware other than passive tags, and negligible adaptations to the underlying RFID-based stock taking process.
The only overhead introduced by our approach is the onetime effort of placing and mapping reference tags to fixtures in a store.

\paragraph{Contributions}
This paper is an extension of our previous work~\cite{woelbitsch2020} in which we first presented a methodology for article-to-fixture predictions, solely relying on read events of passive tags collected using a single handheld reader.
Second, we evaluated our proposed approach in a controlled laboratory environment, as well as in two real-world fashion retail stores, where we achieve article-to-fixture prediction accuracies of more than 90\%.
Third, we published our large-scale data set\footnote{\url{https://github.com/detegoDS/show_me_the_money_dataset}}.

We extend this work by investigating the major challenges related to the implementation of an article-to-fixture prediction system in real-world fashion stores. 
We find that store environments frequently change with respect to article assortments, fixture composition and placement (e.g., removing fixtures from the sales floor).
Therefore, keeping track of fixtures is a complex task, that potentially exceeds reasonable efforts for store staff.
To reduce the effort of managing fixtures, we map articles to store zones (i.e., distinct areas of the sales floor, such as the clearance area) using reference tags that are placed on perpetual positions in a zone instead of movable fixtures.

\section{Related Work}

Kerfoot et al.~\cite{kerfoot2003} conduct a study on how visual merchandising (e.g., the placement and presentation of articles on the sales floor) affects the shopping behavior of customers in retail stores.
They find that factors, such as the fixture design and materials they are made of can influence customers' willingness to buy.
Lea-Greenwood~\cite{lea1998} discusses how modern technology and tools help fashion retailers to design and implement visual merchandising strategies across their stores. 
Moreover, Rizzi and Volpi~\cite{rizzi2017} outline the impact of RFID-enabled visual merchandising and how it can affect sales with respect to the occupied shelf space.
However, they mainly focus on the financial impact and how it can be measured, while our work focuses on the RFID-based implementation that enables the automated assignment of goods to fixtures.

Other RFID-based localization approaches include work from Want and Katabi~\cite{wang2013}, who propose \textit{PinIt}, an RFID-based approach to determine the position of goods based on a synthetic aperture radar generated by a moving antenna.
They compile multipath profiles based on the collected data and use dynamic time warping to approximate locations of tags.
Luo and Shin~\cite{luo2019} introduce \textit{FINDS}, a framework to detect misplaced tags in smart shelf environments.
They use static RFID antennas that measure the phase of tags and improve their measurements using stochastic optimization methods and density-based clustering (i.e., DBSCAN) to detect outliers (i.e., misplaced tags).
Hasler et al.~\cite{hasler2019} use signal strength data collected using a handheld RFID reader to estimate the relative distances between tags by transforming read events into a two-dimensional space using multidimensional scaling.
This enables them to identify misplaced tags by inspecting their neighborhoods.

Moreover, Liu et al.~\cite{liu2015} propose \textit{TagBooth}, which is a method to mine shopping data (i.e., interactions of customers with goods) based on RFID read events.
Specifically, they leverage recorded signal strength and phase information to infer customer actions (e.g., picking up articles).
Similarly, Zhao et al.~\cite{zhao2018} determine interesting, correlated, and popular articles based on signal strength patterns.
Furthermore, they are able to identify hot zones in stores that are frequently visited by customers.
\textit{ShopMiner}~\cite{shangguan2015} and \textit{CBID}~\cite{han2015} are other frameworks designed with the goal to obtain a better understanding of customer behavior in RFID-equipped retail environments.

Other customer-facing applications based on RFID technology in retail stores include digital personal shopping assistants~\cite{ngai2008}, and smart fitting rooms that provide additional information about the articles a customer brings into the dressing room~\cite{al2011}.
These technologies also enable retailers to provide product recommendations for their customers in brick-and-mortar stores~\cite{wolbitsch2019_2}.
Furthermore, RFID-based data-driven methods can be used, for example, to identify counterfeit articles in the supply chain~\cite{staake2005}, or to determine articles that are frequently missed during stocktakes~\cite{wolbitsch2019_1}.

\section{Stocktake Routine \& Practical Impact}
\label{sec:stocktakes}

Stock taking refers to the process of recording and counting all items located in a store (i.e., the inventory).
Using RFID technology, the inventory of a store, which usually consists of several thousand items, can be recorded within a few minutes, while traditional manual stock taking methods can take several days, and often require the temporary closing of the store.

RFID-based stocktakes are usually performed on a daily basis in the morning, right before the store opens.
Store staff use an off-the-shelf handheld UHF RFID reader to record items, which are tagged with passive tags.
To that end, a staff member typically walks alongside the fixtures in an undetermined path and scans the items at short but varying distances. 
In general, RFID-based stocktakes allow retailers to maintain stock accuracies (i.e., the difference between the expected and actual stock) of well beyond 90\%, allowing them to know exactly which items are available in a given store.
This builds the foundation for many \mbox{state-of-the-art} retail technologies, such as \textit{Click\&Collect}, where customers order online and collect the purchased articles in a store.

In this paper, we set out to further leverage the data that is already collected during daily RFID-based stocktakes to localize articles in a store without changing the underlying process, nor requiring extensive RFID parameter tuning by store staff.
Specifically, we use the collected handheld reader data (i.e., timestamp and signal strength of scanned tags) to determine the placement of articles on fixtures.
When enriching this information with other data streams, such as sales data, we can automatically identify ``hot- and cold-fixtures'', simply relying on the existing and unmodified daily stocktake routine, which is a powerful tool to monitor visual merchandising strategies.
Further, additional use cases, such as the compilation of smart pick lists (i.e., lists of articles that should be collected, ordered by their location in the store), can also be calculated based on the presented methodology, which we leave open for future work.

\section{Methodology}

In the first step of our methodology we place reference tags on fixtures.
Next, we perform regular RFID-based stocktakes (see Section~\ref{sec:stocktakes}) using a handheld reader, where we typically collect
multiple read events per RFID tag.
A read event consists of the Received Signal Strength Indication (RSSI), the timestamp of the read event (millisecond resolution), and the unique identifier encoded in the tag.
Hence, we obtain a time-ordered sequence of RSSI measurements for all tags (i.e., reference tags and item tags), which we aggregate on article and fixture level.
Next, we approximate distances between articles and fixtures using two different approaches, and finally transform distances into probabilities, which we use to predict article-to-fixture assignments.
Moreover, we extend our approaches to leverage historic stocktake data to further improve prediction performance over time.

\subsection{Fixture Tagging}

For our work we enumerate all fixtures and place reference tags on them.
Note that the definition of what constitutes a fixture in a store setting is fluid and depends on local circumstances.
For example, multiple smaller adjacent fixtures that are individually tagged, can be aggregated to form one larger ``logical'' fixture. 
We achieve this by combining all reference tags of the corresponding fixtures.
Such a ``logical'' fixture can then be used to identify a ``zone'' in a store, such as the clearance area or the area containing the latest articles.
Ultimately, the store layout in combination with the required article-to-fixture mapping granularity, defines the trade-off between keeping the number of fixtures manageable (i.e., managing the mapping between fixtures and tags), while still providing enough information about article locations to generate valuable insights for the retailer. 

Additionally, RFID technology itself imposes limits on the potential granularity of fixtures, due to the noisy nature of the reading results.
For example, small fixtures located very close to each other are potentially very difficult to distinguish.

\subsection{Read Event Aggregation}

\begin{figure}[!t]
    \centering
    \includegraphics[width=0.75\textwidth]{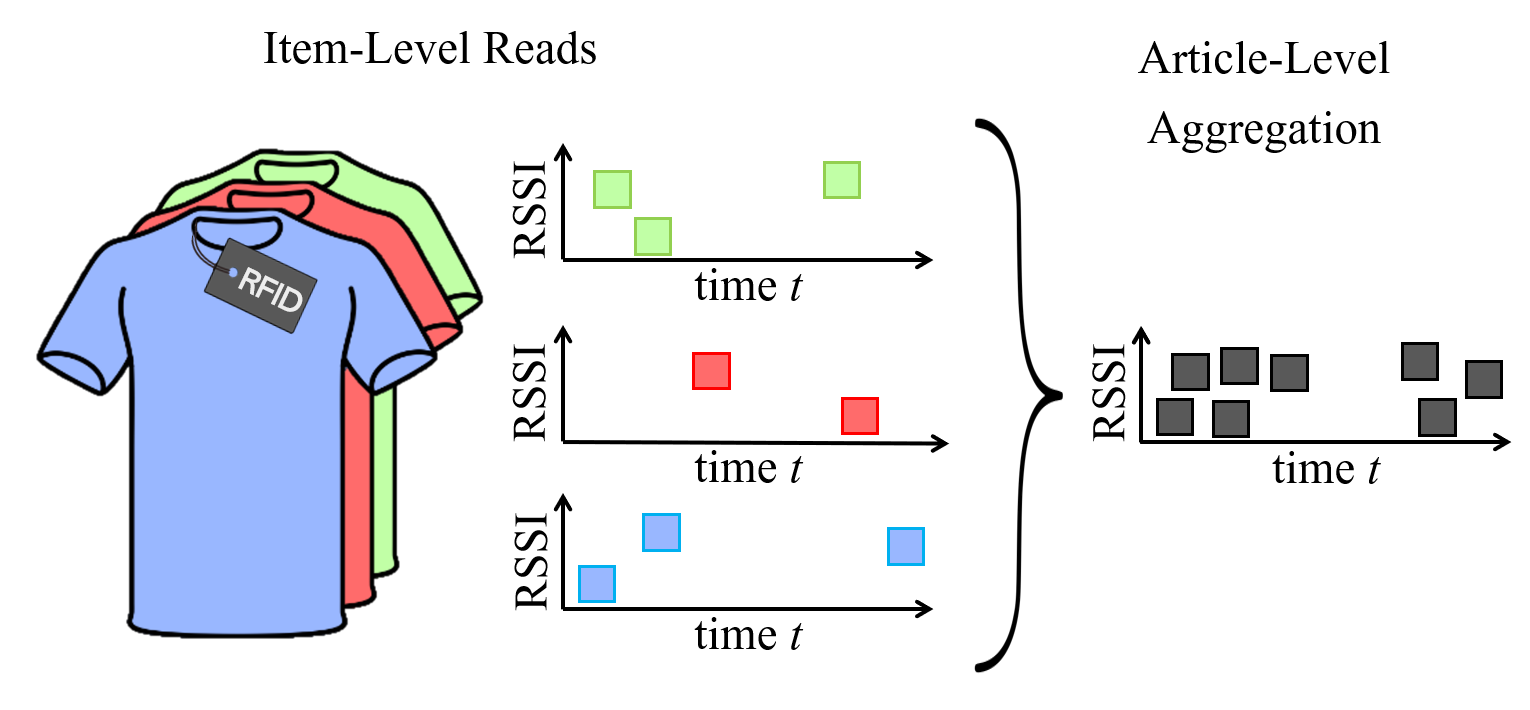}
    \caption{\textbf{Read Event Aggregation.} As all items on an ``article stack'' belong to the same article (invariant with respect to color and size) we can aggregate the individual read events to get a more robust signal that can be used to predict the fixture on which the ``article stack'' is located.}
    \label{fig:read-aggregation}
\end{figure}

Fashion retailers usually place multiple items of an article on the same fixture so that customers have a selection of different sizes in one location (i.e., ``article stacks'').
As the number of read events per individual item are limited, and the recorded signal strength might be noisy, we leverage the fact that related items are already physically close to each other.
Specifically, we perform an aggregation step where read events of items belonging to the same article are combined, thus allowing us to obtain a more dense series of reads on article-levels (see Figure~\ref{fig:read-aggregation}).
Similarly, reference tag reads, which are located on the same fixture, are also aggregated to achieve the same effect.

\subsection{Distance Approximation Approaches} 

We have implemented two different approaches to tackle the problem of identifying the most probable fixture for an article. 
Specifically, we calculate the ``distance'' between each article read event sequence and all other sequences that represent fixtures.
Note that distance is an arbitrary measure that depends on the used fixture prediction approach.

\paragraph{Parameter Estimation}
Both presented approaches depend on a set of parameters that we obtain using hyperparameter optimization, which is a commonly performed practice in the machine learning domain.
To that end, we conduct 5 stocktakes, each with the RFID reader configured in Session~0 and in Session~1 in a laboratory setting, where locations of articles are known beforehand.
We then apply our approaches to the collected stocktake data using a wide range of different parameter value combinations, to determine the best performing configuration.
Note that we use the selected parameters for all our controlled (real-world) experiments to evaluate their applicability in practical scenarios, where extensive parameter tuning is significantly more costly.

\paragraph{DBSCAN-based Approach}
Our first approach is based on Density-based Spatial Clustering of Applications with Noise (DBSCAN)~\cite{ester1996}. 
Essentially, with DBSCAN, we try to identify the shortest distance in time between clusters of read events for articles and fixtures.
Specifically, DBSCAN groups timestamps of reads, which are close to each other based on Euclidean distance and a minimum number of points in their neighborhood, allowing us to identify one or more clusters per read event sequence, while simultaneously filtering noise.

For this algorithm we perform MinMax scaling on the time stamps of the read events such that the first read event happens at \( t = 0 \) and the last one at \( t = 1 \).
Furthermore, we perform MinMax scaling on the signal strengths and remove all read events with a low RSSI value before clustering to reduce the influence of noisy reads.
Specifically, we remove read events with an RSSI value smaller than the 0.8 quantile for Session~0 and the 0.77 quantile for Session~1. 

For DBSCAN, $\epsilon$ defines the maximum Euclidean distance between two read events to be considered neighbors.
We set this parameter to \( \epsilon = 0.085 \) for Session~0 and \( \epsilon = 0.068 \) for Session~1.
Further, we define that a dense region (i.e., cluster) must contain at least 8 reads for Session~0 and 7 for Session~1, which is the second important parameter for DBSCAN. 

\begin{figure}[!t]
    \centering
    \includegraphics[width=0.65\textwidth]{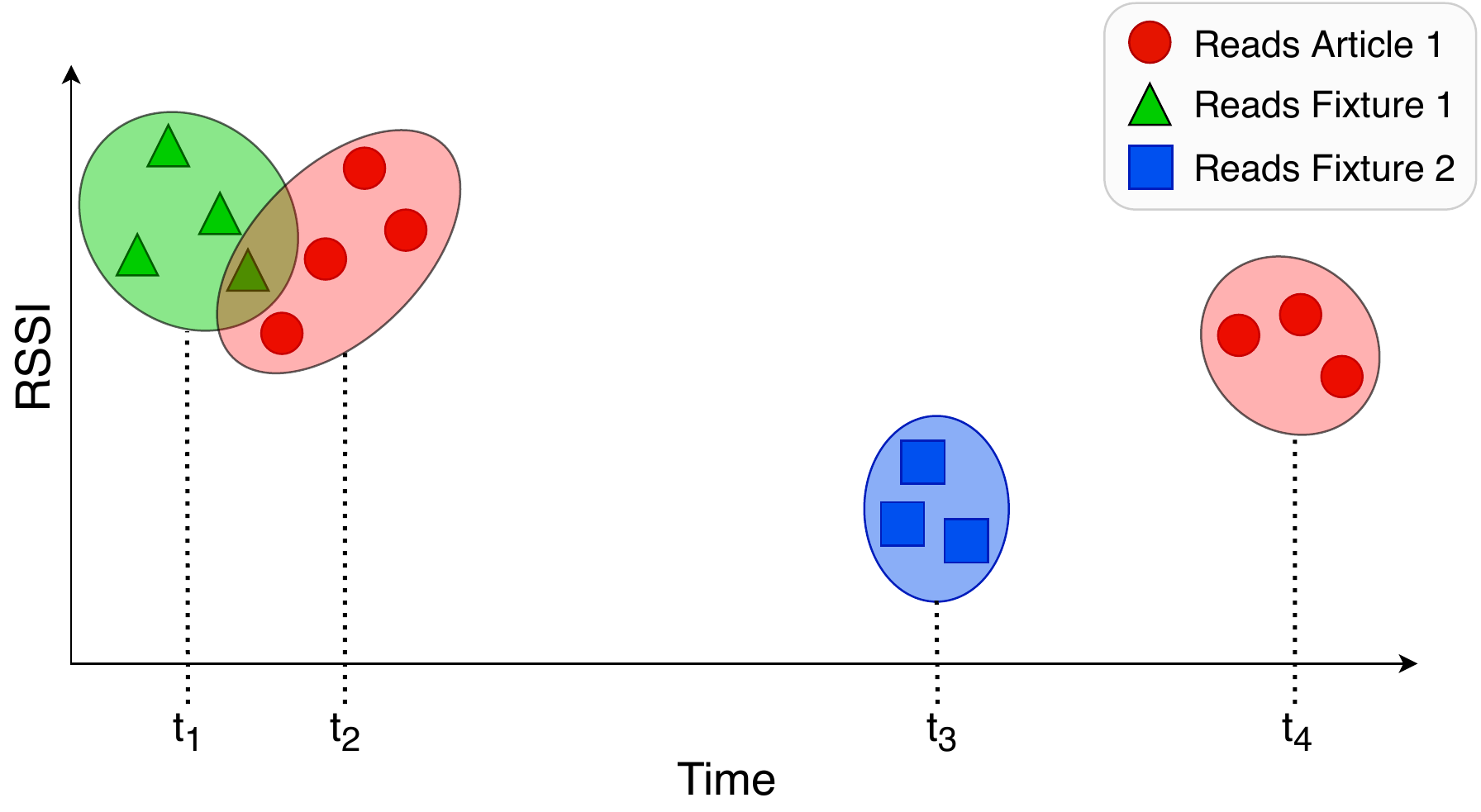}
    \caption{\textbf{DBSCAN-based Approach Illustration.} We show clusters of read events obtained by DBSCAN for an article (\protect\mycircle{red}) and two fixtures (\protect\mytriangle{green}, \protect\mysquare{blue}) (i.e., reference tags). We use the minimum time difference between cluster centroids belonging to the fixtures and the article to determine the distance between them. For example, the distance between Article~1 and Fixture~2 is \( d_{1,2} = \mathopen| t_4 - t_3 \mathclose| \), as \( \mathopen| t_4 - t_3 \mathclose| < \mathopen| t_3 - t_2 \mathclose| \).
    However, we ultimately assign Article~1 to Fixture~1 in this example based on the overall smallest distance  \( d_{1,1} = \mathopen| t_2 - t_1 \mathclose| \).}
    \label{fig:dbscan}
\end{figure}

Finally, the distance between a read event sequence of an article and a fixture is the minimum distance in time between the respective cluster centroids (see Figure~\ref{fig:dbscan}).

\paragraph{DTW-based Approach}
Our second approach is based on Dynamic Time Warping (DTW)~\cite{kruskal1999}, which can be used to determine the distance between two time series with different characteristics.
Therefore, the first step is to convert the irregular sequence of read events, consisting of timestamps and measured signal strengths (i.e., RSSI values), to a time series.
To that end, we resample the read events for articles and fixtures to obtain equally spaced series of RSSI values that start and end at the same time.
For our work, we resample read events with a resolution of 0.2 seconds for Session~0 and 0.1 for Session~1 and sum the RSSI values within a time window.

Note that due to the definition of the RSSI, the corresponding values in our setup are in the range of \( -80 \) to \( -20 \) dBm.
To simplify our calculations, we transform the RSSI values to positive numbers by adding 100 to the original values.
This also conveniently allows us to set the signal strength in the resampled time series to 0 whenever no read event occurred within a time window.
Furthermore, we discard read events with an RSSI value smaller than the 0.4 quantile for Session~0 and 0.5 quantile for Session~1 to filter noise.

We then use DTW to determine distances between article and fixture time series.
To that end, DTW maps each point in the article time series to one or more points in the fixture time series.
The actual distance between two time series is determined by the cost originating from this matching process.
Note that for DTW we set the maximum window size \( w \), which is the maximum shift allowed to match two points, to \( w = 9 \) for Session~0 and \( w = 12 \) for Session~1.

\subsection{Article-to-Fixture Prediction}

Next, we assign articles to fixtures based on distances approximated using the previously discussed approaches.
To that end, we transform distances into probabilities via inverse distance weighting.
More formally, the probability \( p_{i,j} \) that article \( i \) is located on fixture \( j \) is given by
\begin{equation}
  p_{i,j} = \frac{\frac{1}{d_{i,j}^2}}{\sum_{k}^{N}{\frac{1}{d_{i,k}^2}}},
\end{equation}
where \( d_{i,j} \) is the distance between article \( i \) and fixture \( j \) and \( N \) is the total number of fixtures.
Note that we use squared distances to penalize larger distances obtained by our approaches.
Finally, we assign article \( i \) to the fixture with the largest probability.

\subsection{Leveraging Historic Information}

The quality of the collected data does not only depend on the underlying RFID technology, but also on how thorough stocktakes are performed.
As a result, recorded stocktake data may contain noise, a limited number of read events, or other characteristics that can have a negative impact on our results. 

To tackle this issue, we leverage historic information from previous stocktakes.
Due to the fact that article placements on fixtures are relatively stable (i.e., extensive rearrangements typically only happen when new fashion lines are introduced or seasons change) and RFID-based stocktakes are performed frequently (i.e., daily), we can assume that whenever we are confident about a fixture assignment for an article (i.e., high probability for the fixture), the article will most likely also be at the same fixture after the next stocktake.
Hence, to compensate for stocktakes that contain many noisy reads we propose a Bayesian approach in which we consider evidence from previous assignments when updating fixture assignment probabilities.

To that end, we calculate a certainty measure of the fixture assignment for each article.
We define certainty as \(1 - H/H_{\max}\), where \( H \) denotes the entropy of the fixture probability distribution, and \( H_{\max} = \log_2N\) is the maximum entropy (i.e., the entropy of the uniform distribution).
As a result, if all fixtures are equally likely, the certainty of the assignment would be $0$, while a distribution with an assignment probability for only a single fixture would lead to a certainty of $1$.
We scale the probability distributions of the current and the previous stocktake by their respective certainty scores and combine the two distributions by adding the individual fixture probabilities.
Finally, we normalize the newly created distribution such that the entries sum up to 1 and assign the article to the fixture with the highest probability.

\subsection{Performance Evaluation Metrics}

We evaluate our proposed methodology using two different metrics.
First, we measure the accuracy of our computed assignments (i.e., the fraction of correctly assigned articles to fixtures).
Second, as fixtures are often located very close to each other, we are also interested in how accurate our models are with respect to the distance to the actual fixture. 
For example, based on the local properties in stores, we expect that some articles will be be assigned to neighboring fixtures (i.e., ``off-by-one'' errors).
To verify this intuition, we compile a distance matrix for the fixtures based on a regular two-dimensional grid that we laid over the floor plan.
We use the Chebyshev distance metric as distance measure between two fixtures on the grid as it is easy to interpret.
For example, a distance error of 1 using this metric corresponds to the incorrect assignment to a fixture on the neighboring square on the grid, independent of the direction.

\section{Experimental Setup}
To evaluate and test our article-to-fixture prediction approaches, we perform multiple experiments in three different settings.
First, we conduct a controlled laboratory experiment to determine the feasibility of our methods.
Next, to evaluate the performance of our methodology in a real-world scenario, we visit two brick-and-mortar stores of a large international fashion retailer and conduct additional experiments. 
Finally, we investigate the challenges related to the integration of an article-to-fixture prediction system in the day-to-day business of fashion retail stores within the scope of a first pilot.
Specifically, this pilot consists of a permanent installation of reference tags, a granular fixture management, and unsupervised stocktakes performed by store staff.

\subsection{RFID Setup}

For all experiments we use a single off-the-shelf RFID handheld reader (i.e., Zebra RFD8500 or Bluebird RFR900) and passive RFID tags that are commonly used in the fashion retail domain.
During the pilot project we use ``on-metal'' tags as reference tags (see Figure~\ref{fig:on-metal-tag}). 
These tags can be placed on fixtures that are made of metal without compromising the readability of the tags.
Note that, in general, all tags that we use are comparable in terms of readability.

We configure the reader to use Session~0 in \textit{AB flip} mode to collect as many read events as possible.
However, we also conduct stocktakes with Session~1, which reduces the number of reads per tag, as in contrast to Session~0, the tags have an added cool-down period before responding again. 
This allows for higher throughput during stocktakes, but increases the difficulty of predicting fixtures for articles.
Note that stocktakes performed during the pilot are done using Session~1.

Furthermore, we set the power level of the reader to its maximum in all environments (30 dBm; corresponding to 1W) and connect the RFID handheld device via Bluetooth to an Android smartphone, running the software that collects and stores the read events from the RFID reader.

\subsection{Controlled Lab Experiments}

\begin{figure}[!t]
  \center
  \subfloat[Schematic of Laboratory Setup\label{fig:lab-setup:schematic}]{%
   \includegraphics[width=0.5\textwidth]{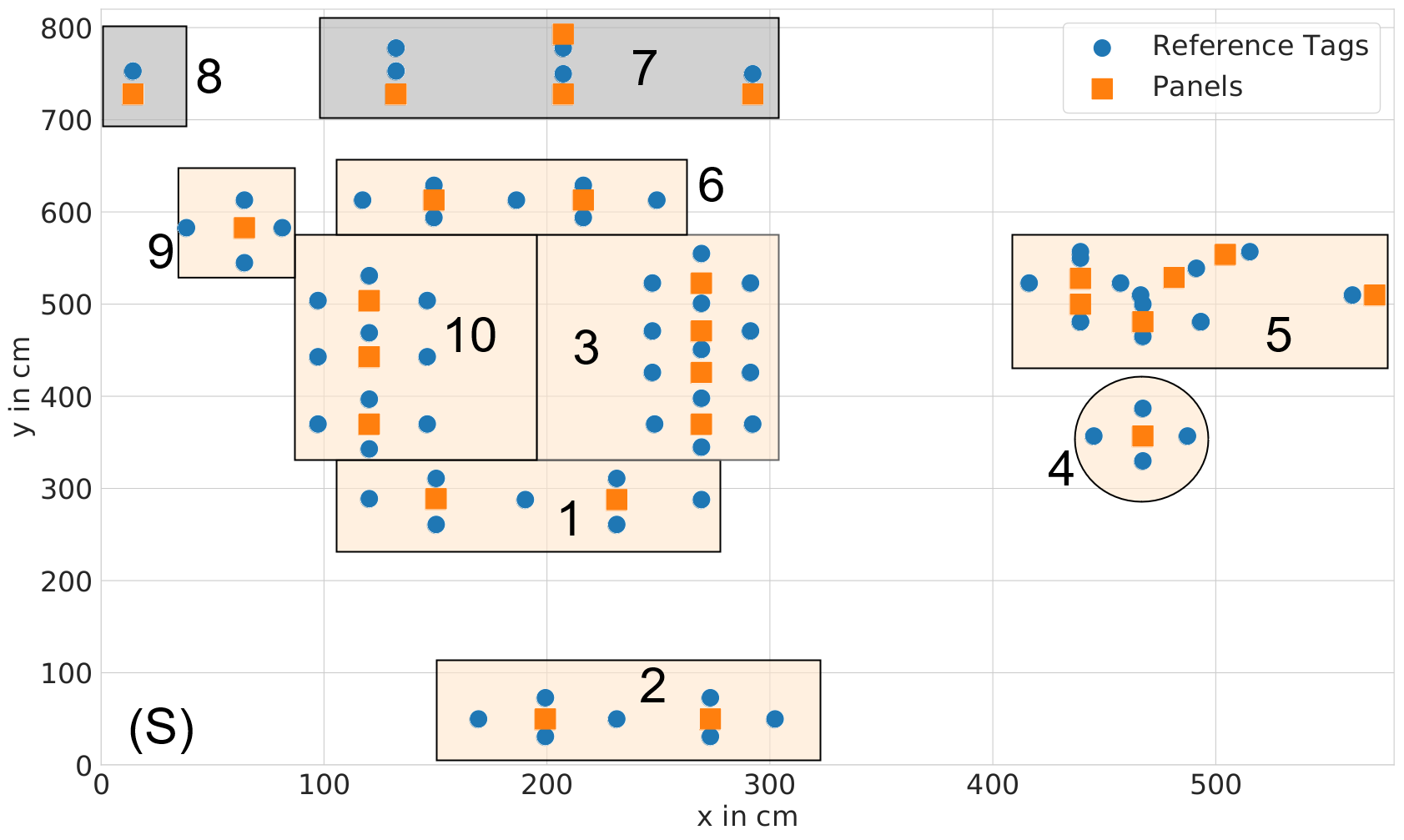}
 }
 \hspace{2em}
 \subfloat[View from (S) in Schematic\label{fig:lab-setup:setup-3}]{%
   \includegraphics[width=0.4\textwidth]{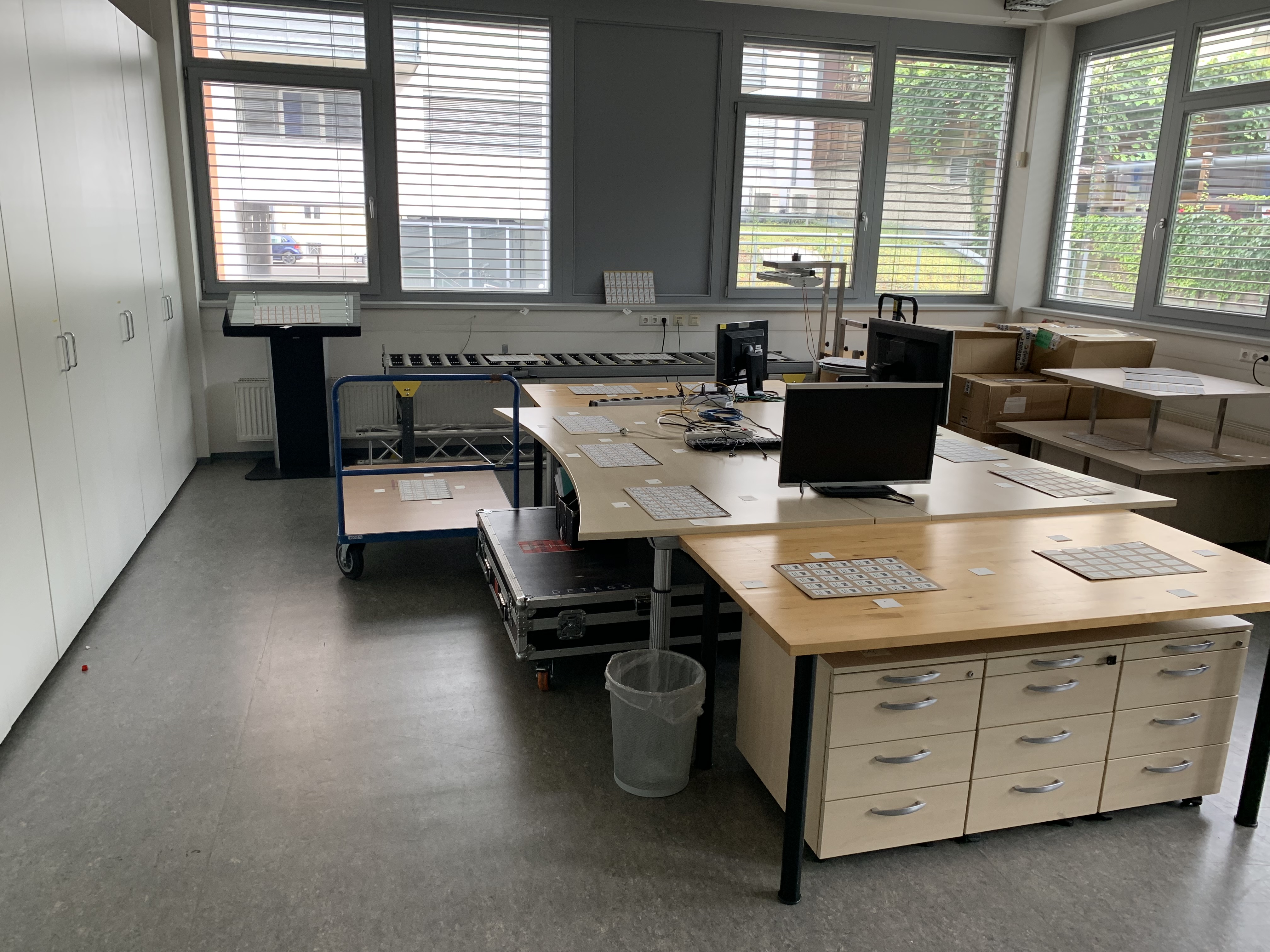}
 }

 \caption{\textbf{Controlled Lab Setup.} Rectangles and squares represent fixtures (enumerated by the numbers in Figure~\ref{fig:lab-setup:schematic}). Orange markers represent panels (i.e., ``article stacks'') and blue markers individual reference tags. Note that panels and reference tags on Fixture 5 are overlapping in our schematic, as it is a multi-layered table (see top right of Figure~\ref{fig:lab-setup:setup-3}). In total, we distributed 27 panels (each with 24 to 36 tags) and 74 reference tags across 10 fixtures. (S) marks the starting point for all stocktakes.
 }
 \label{fig:evaluation-steps}
\end{figure}

We conduct our first experiment in a controlled laboratory setting (see Figure~\ref{fig:evaluation-steps}), where fixtures are located in a large room that also contains worktables, toolboxes, racks, and a conveyor belt.
Specifically, we use a total of 10 fixtures, which consist of wooden tables (fixtures 1 to 6, and 10), a conveyor belt (fixture 7) as well as a standing desk touch-screen (fixture 8) and an open cart (fixture 9).

We place a total of 27 carton panels on the fixtures, each containing 24 to 36 printed RFID labels in a regular pattern.
These panels simulate items belonging to the same article, mimicking commonly used ``article stacks'' in multiple sizes and colors on the same fixture.
In total, we distribute 916 items and 74 reference tags across 10 fixtures (see Figure~\ref{fig:lab-setup:schematic}).
Note that we use similar RFID labels for reference tags and that we manually select the required number of reference tags placed on a fixture based on its size and location.

We conduct a total of 14 stocktakes in Session~0 and 19 in Session~1, all using the same starting position (see (S) in Figure~\ref{fig:lab-setup:schematic}). 
However, for each experiment we vary the walking path and walking speed while reading the tags.
Furthermore, we compile a ground truth dataset, which maps both item and reference tags to fixtures in this setup.
We use this ground truth to evaluate the performance of our proposed methods on the performed stocktakes.

\begin{table}[b!]\footnotesize
\center
\caption{\textbf{Controlled Experiments Data Set Properties.} For each controlled experiment we list the number of articles and items, as well as the number of reference tags used in the setup.
Furthermore, we state the number of fixtures.
Note that not all fixtures were equipped with the same number of reference tags.
Finally, we list the number of conducted stocktakes and their average duration in parenthesis. The figures for the real-world pilot are not listed in this table as they vary over time.}
\label{tab:experiment-dataset}
\begin{tabular}{lcccc} \toprule
Experiment & Articles (Items) & Ref. Tags & Fixtures & Stocktakes\\ \midrule
Lab & 27 (916) & 74 & 10 & 33 (01:43) \\ 
Store~A & 197 (1,739) & 156 & 33 & 11 (07:29) \\
Store~B & 200 (1,977) & 98 & 23 & 11 (10:06) \\ \bottomrule
\end{tabular}
\end{table}

\subsection{Controlled Store Experiments}
\label{subsec:controlled-store-exp}

\begin{figure}[!t]
  \center
  \subfloat[Schematic Store~A\label{fig:store-setup:schematic}]{%
   \includegraphics[width=0.7\textwidth]{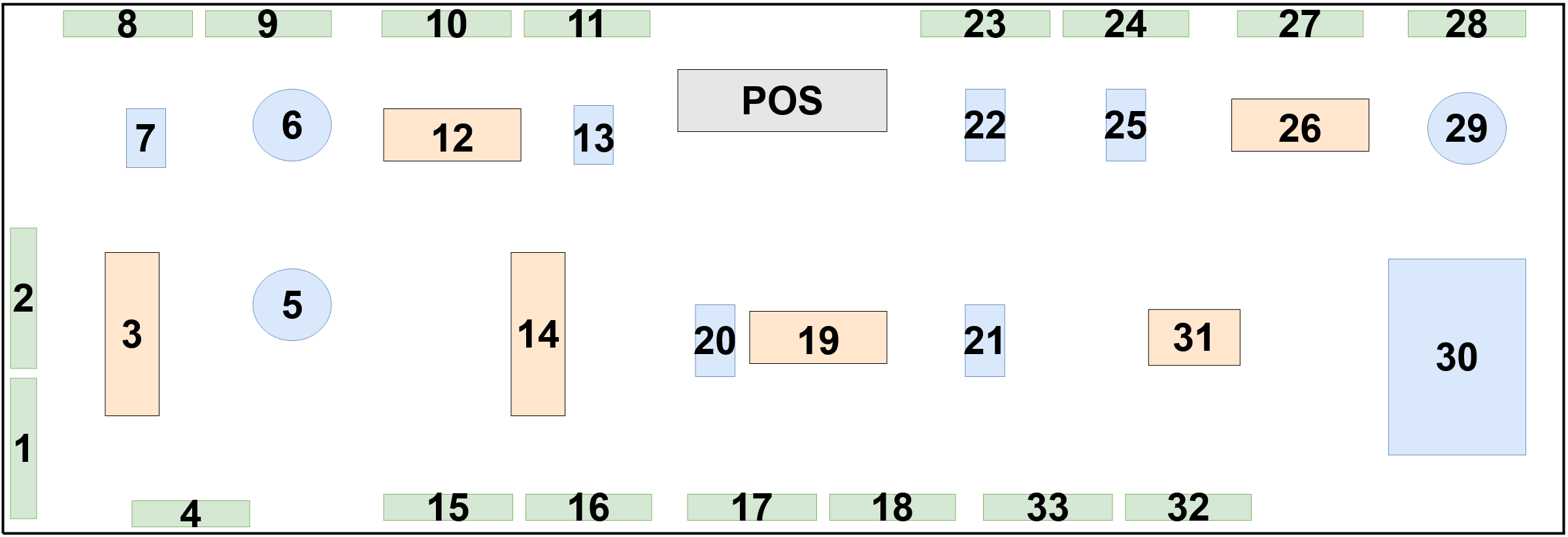}
 }
 \hspace{2em}
 \subfloat[Ref. Tag Placement\label{fig:tag-placement}]{%
   \includegraphics[width=0.17\textwidth]{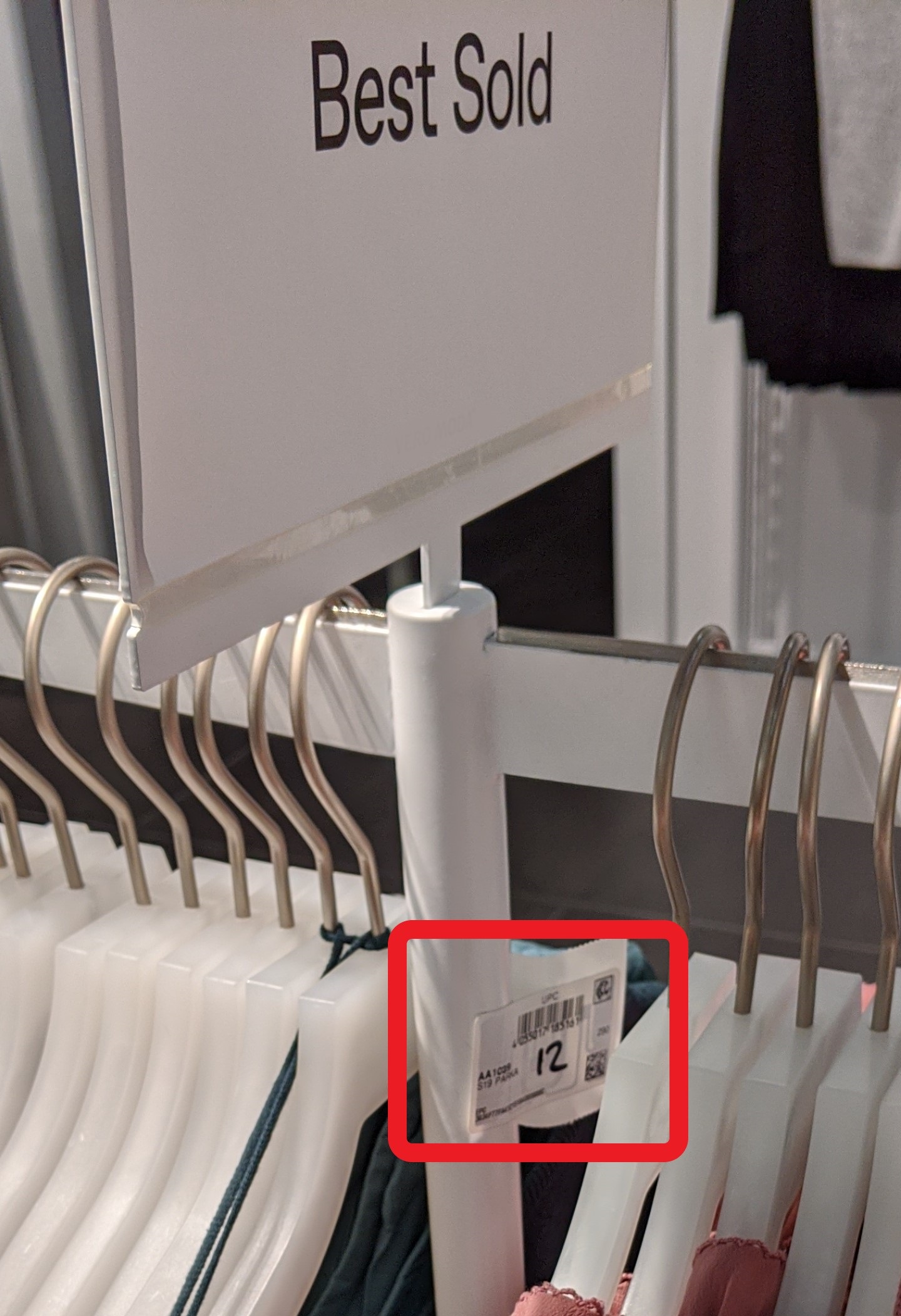}
 }

 \caption{\textbf{Controlled Store Setup.} Rectangles and circles on the schematics in Figure~\ref{fig:store-setup:schematic} depict fixtures located on the sales floor of Store~A, which has a size of about 180 square meters. Overall there are 33 fixtures located on the sales floor, which are either tables (brown), shelves (green), or clothing racks (blue). Furthermore, the point of sale (POS) is also shown. Figure~\ref{fig:tag-placement} shows an example for the placement of a reference tag on a fixture in this store during the controlled experiments.}
 \label{fig:store}
\end{figure}

In addition to the stocktakes we perform in the lab, we also conduct stocktakes on the sales floors of two brick-and-mortar stores of an international fashion retailer (see Figure~\ref{fig:store}).
We select the two stores based on their different inventory sizes, shop areas, number of fixtures, and the materials fixtures are made of.
Store~A is equipped with several fixtures that are made of metal, which can negatively affect reading performance during stocktakes due to signal reflections, while Store~B is mainly equipped with wooden fixtures.
The fixtures in the stores cover tables (on which items were placed in stacks), shelves (that can also include clothing rails), and clothing racks (that often contain a large quantity of individual articles).

Overall we observe 197 different articles (1,739 items) on 33 fixtures in Store~A, and 200 different articles (1,977 items) on 23 fixtures in Store~B.
Despite fewer items on the sales floor, Store~A has a larger number of items per square meter and more fixtures compared to the Store~B (cf. Table~\ref{tab:experiment-dataset}).
All items in the stores were already equipped with RFID tags during manufacturing. 

Similarly to the laboratory setup, we use printed RFID labels as reference tags and determine the number of reference tags per fixture according to fixture sizes and types.
In general, larger fixtures are tagged with more reference tags than smaller ones.
For fixtures with metallic rails we place reference tags on plastic hangers, which we evenly distribute along the rails, as placing them directly on metal surfaces would negatively affect their readability. 

In total, we conduct 11 stocktakes (8 in Session~0, 3 in Session~1) in both stores.
Note that unlike in the laboratory setting, we do not use a fixed starting position for the stocktakes and heavily vary walking paths to mimic realistic stocktakes as close as possible.
We also compile a ground truth for this setting, so that we can reliably determine the accuracy of the proposed method based on the performed stocktakes in both stores.

\subsection{Real-World Pilot}

The previous experiments are not designed to fully comply with all real-world requirements that are entailed when introducing an article-to-fixture mapping feature in brick-and-mortar retail stores.
For example, placing printed RFID labels on plastic hangers as reference tags (see controlled store experiments in Section~\ref{subsec:controlled-store-exp}) is not a viable option, as reference tags must be permanently attached to fixtures.
Therefore, in this unsupervised pilot we asses the practicability of our article-to-fixture mapping feature, and implications for a potential mass roll-out.

\begin{figure}[!t]
    \centering
    \includegraphics[width=0.65\textwidth]{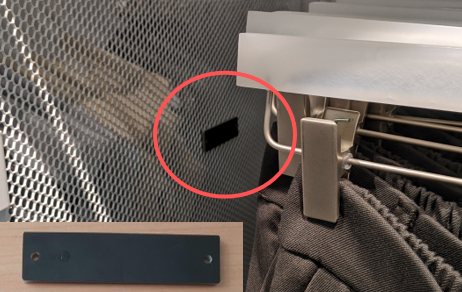}
    \caption{\textbf{On-metal RFID Tag.} For our real-world pilot we place special ``on-metal'' RFID tags on the surfaces of metal fixtures (e.g., on the metallic mesh as depicted) as reference tags. Note, to prevent the introduction of too many different types of tags we use the same ``on-metal'' tags as reference tags for all fixtures, independent of the material they are made of.}
    \label{fig:on-metal-tag}
\end{figure}

We conduct this pilot in the same two brick-and-mortar stores over multiple weeks.
First, we perform an enumeration of fixtures and place reference tags in both stores.
To that end we use ``on-metal'' tags (see Figure~\ref{fig:on-metal-tag}) that are placed directly on fixtures and can stay there for extended periods of time, without disrupting daily operations.
Once a week, we generate a ground truth article-to-fixture mapping, and perform stocktakes on the sales floors in addition to the regular stocktakes that store staff performs during normal operation on that day.
Therefore, we obtain weekly mapping accuracies for the stocktakes performed by us and store staff.

For the pilot we enumerate a total of 106 fixtures in Store~A and 74 fixtures in Store~B, which is significantly more granular than in our controlled experiments (cf. Table~\ref{tab:experiment-dataset}) as our setup needs to account for (parts of) fixtures that might be rearranged. 
For example, a two-layered fixture, compiled of two tables with different heights that can be detached from each other, is counted as two different fixtures, as they could be used as individual fixtures in a different setup.
For our algorithms to work, such granular fixture management is required 
in order to perform accurate reference tag aggregation. 
It is also possible that individual parts, or entire fixtures, are temporarily moved from the sales floor to storage or vice versa.
Therefore, we also enumerate fixtures that are currently not in use.

Our proposed article-to-fixture prediction methods assume that ``article stacks'' of the same article are only represented on one fixture. 
However, in our real-world pilot, we find that this assumption does not always hold.
Specifically, we observe that the same article in different colors can be found on separate fixtures  (e.g., a particular T-shirt in multiple sizes is placed in red on fixture 1 and in blue on fixture 2).
Therefore, for the pilot project we also consider color information when aggregating read events (i.e., aggregating read events of the same article in one specific color but different sizes).
This then allows us to map the same article in different colors to different fixtures.
Hence, we also increase the number of articles that have to be mapped accordingly, which is a more challenging task, but at the same time enables the retailer to be more flexible with article placement on the sales floor.

\section{Results \& Discussion}

\subsection{Controlled Lab Environment}

First, we evaluate our proposed methods to predict article-to-fixture assignments in a controlled laboratory setup (cf. Lab in Table~\ref{tab:results}).
In this setting, we place 26 different articles (i.e., panels containing RFID tags) on 10 fixtures, that are located within a few meters.
Assigning articles to fixtures in this scenario at random leads to an accuracy of about 10\%, which represents a baseline for our experiments.
Our DBSCAN-based approach outperforms this baseline with more than 78\% accuracy on average for stocktakes conducted in both Session~0 and Session~1.
We find even better results using our DTW-based approach, which achieves an average accuracy of 89.9\% for Session~0 and 90.9\% for Session~1.

When leveraging historic information (i.e., article-to-fixture assignments computed in previous stocktakes) we find an improvement of accuracy for both our approaches.
In the case of our DBSCAN-based method by 9.9\% (7.1\%) and for DTW-based method by 2.6\% (2.2\%) for Session~0 (Session~1).
Note that we do not evaluate our results in this setup with respect to the distance for wrong article-to-fixture assignments due to the small number of fixtures.

\begin{table*}[!b]

\centering
\scriptsize

\caption{\textbf{Experimental Results in Controlled Environments.} In this table we report the accuracy (i.e., the percentage of correctly assigned articles to fixtures) and distance to the correct fixture for wrong article-to-fixture predictions (i.e., the error). The distance error indicates how close the predicted fixture is to the correct one on a regular grid laid over the area in which the fixtures are located. We report averages (and standard deviation) for both metrics.
}
\label{tab:results}

\begin{tabular}{@{}clcccccccc@{}}
\toprule
\multirow{2}{*}{\textbf{Session}} & \multicolumn{1}{c}{\multirow{2}{*}{\textbf{Experiment}}} & \multicolumn{2}{c}{\textbf{DBSCAN}} & \multicolumn{2}{c}{\textbf{DBSCAN w/ History}} & \multicolumn{2}{c}{\textbf{DTW}} & \multicolumn{2}{c}{\textbf{DTW w/ History}} \\
 & \multicolumn{1}{c}{} & Accuracy {[}\%{]} & Error & Accuracy {[}\%{]} & Error & Accuracy {[}\%{]} & Error & Accuracy {[}\%{]} & Error \\ \midrule
\multirow{3}{*}{0} & Lab & 78.60 (17.84) & - & 88.48 (6.79) & - & 88.89 (6.78) & - & 91.45 (3.74) & - \\
 & Store~A & 81.51 (5.57) & 1.16 (0.09) & 88.34 (6.92) & 1.21 (0.20) & 70.51 (4.54) & 2.03 (0.32) & 71.72 (4.54) & 2.62 (0.94) \\
 & Store~B & 88.31 (6.51) & 1.65 (0.46) & 93.18 (3.65) & 1.75 (0.52) & 80.82 (5.36) & 2.24 (0.70) & 83.89 (2.43) & 2.13 (0.75) \\ \midrule
\multirow{3}{*}{1} & Lab & 78.21 (12.20) & - & 85.36 (8.99) & - & 90.86 (6.39) & - & 93.09 (4.41) & - \\
 & Store~A & 74.89 (5.54) & 1.46 (0.54) & 78.51 (2.31) & 1.69 (0.53) & 65.74 (3.97) & 3.19 (1.51) & 64.04 (2.37) & 4.57 (0.19) \\
 & Store~B & 78.44 (7.83) & 1.74 (0.10) & 83.63 (12.57) & 1.74 (0.17) & 74.85 (5.39) & 1.84 (0.13) & 74.65 (5.09) & 1.92 (0.02) \\ \bottomrule
\end{tabular}

\end{table*}

\paragraph{Discussion}
While the laboratory setting provides stable conditions for our experiments, the setup differs from a real-world environment in several dimensions (e.g., in terms of the number of articles or fixture types).
However, it also exhibits realistic properties, such as reflecting and absorbing surfaces, and representative distances between fixtures.
Hence, the high accuracies we achieve for both our approaches are promising.

Moreover, in the laboratory setting we obtain similar performance for both sessions, although we have on average 28.5\% fewer read events for stocktakes in Session~1 than in Session~0.
This indicates that more read events in such a rudimentary environment have a limited impact on our methods.

When comparing the two approaches directly with each other, we find that the DTW-based approach outperforms the clustering-based approach.
In general, performance for the time-series-based method is also more stable across different experiments, which is evident in the lower standard deviation of accuracy across multiple stocktakes.
However, the observed gain in performance when leveraging historic stocktake information is larger for the DBSCAN-based method, which suggests that this additional input stabilizes prediction results to a larger extent.
We can also observe this in the reduced standard deviation of achieved accuracies across all stocktakes.

\subsection{Controlled Store Environment}

Next, we evaluate our methods in two real-world environments (cf. Store~A and Store~B in Table~\ref{tab:results}).
To that end, we conduct stocktakes on the sales floors of two fashion retail stores with different characteristics.
As the number of fixtures is larger at these sites, the baseline, which randomly assigns articles to fixtures, performs worse compared to the controlled laboratory environment with accuracies between 2\% and 6\%.

In Store~A we achieve a better average accuracy using our clustering-based approach compared to the lab setting in Session~0 with 81.5\%, and 88.3\% when leveraging historic information.
However, for Session~1 stocktakes in Store A, we are not able to match the accuracy achieved in the lab using the same approach with 74.9\% and 78.5\%.
While the DTW-based approach achieves the overall best performance in the lab setting, accuracy is low for stocktakes in Store~A.
For example, accuracy for Session~1 is only 65.7\%, and slightly decreases to 64.0\% when leveraging historic stocktakes. 

When comparing the accuracies of our proposed methods between the two stores, we find that we can, in general, achieve a better performance across all approaches and sessions in Store~B.
Especially the DBSCAN-based approach that uses historic information achieves an average accuracy of 93.2\% for Session~0 stocktakes at this site.

In contrast to the stocktakes we conduct in the laboratory, we also investigate the error with respect to the distance to the correct fixture for wrong article-to-fixture predictions.
We find that these errors are generally very small for the DBSCAN-based approach across all settings (see Table~\ref{tab:results}).
Note that the minimum possible distance error is 1, which corresponds to the assignment of an article to the fixture that is located right next to the correct one.
The average distance error for wrong predictions for stocktakes conducted in Session~0 in Store~A is as low as 1.16 using this approach.
Nevertheless, we also find that especially the DTW-based approach not only struggles in terms of accuracy in Store~A, but also with larger distance errors for wrongly assigned fixtures.

\paragraph{Discussion}
In contrast to the laboratory setting, we find that our time-series-based approach is outperformed by the clustering-based approach in a real-world environment.
The margin between the two methods is also often very large.
For example, for Session~0  stocktakes in Store~B, where we can achieve our overall best average accuracy with more than 93\%, the DTW-based approach only achieves an average accuracy of 83.9\%.
We hypothesize that the differences in performance between the two approaches in the real-world and laboratory setting is related to the number of read events collected during stocktakes.
The number of items that are recorded in stores is much larger, which is beneficial for the clustering-based approach, as more stable and distinct clusters can be formed.
In contrast to our expectations, the DTW-based approach can not take advantage of the additional data.
By investigating different hyperparameter settings based on stocktakes performed in stores, we find that the DTW-based approach is also able to achieve similar performance as the DBSCAN-based approach in both stores.
However, we find that the DTW-based approach requires a more careful selection of its parameters, dependent on the store environment.
In general, a larger resample window and lower threshold for the removal of reads with smaller RSSI values seems to be beneficial for this method.
Nevertheless, for many retailers it is not feasible to generate a ground truth for their stores to fine-tune parameters due to the associated substantial effort that has to be made.
Hence, our goal is to build a model that works for a variety of store setups.
Our DBSCAN-based approach appears to be better suited to tackle this requirement as we can achieve good performance across three different sites.

Moreover, we are inclined to attribute the difference in performance between the stores to two factors.
First, the two store layouts differ substantially.
Store~A has more fixtures located on a smaller sales floor, and at the same time fewer items per fixture, compared to Store~B.
Hence, in Store~A, we have less data to discriminate between a larger number of fixtures.
The two stores also differ in the materials that fixtures are made of.
Many fixtures in Store~A are made of metal, which increases the potential for reflections of RFID signals, and leads to more noisy data.
Second, we also experiment with the placement of reference tags on the fixtures between the two stores.
For Store~A, we place reference tags directly on the fixtures (e.g., on shelf frames), while for Store~B we put reference tags on plastic hangers and wooden surfaces.
Therefore, the direct placement of reference tags on fixtures that are made of metal in Store~A is another potential factor for lower accuracies at this site.
Note that the placement of reference tags on tables was consistent across both stores.
These two factors render Store~A a more challenging environment.
Nevertheless, by leveraging historic article-to-fixture assignments with the DBSCAN-based approach we are still able to achieve an average accuracy of 88.3\% in Store~A, which is a promising result for many real-world applications.

Moreover, the distance error of 1.21 in the same store is encouraging as well, as the majority of wrong article-to-fixture predictions are assignments to one of the neighboring fixtures.
This highlights that our proposed approach is not only able to accurately assign articles to the fixture that they are actually located on, but it is also not far off in terms of distance in case of a wrong prediction.
This is an important trait for many real-world use cases as, for example, store staff can usually find articles that they want to retrieve in the near proximity of the predicted location in the few cases the prediction is wrong.

\subsection{Real-World Pilot}

Finally, we evaluate our proposed method with respect to performance as well as practicability within the scope of a real-world pilot. 
The environment in which the pilot takes place is the same as in the controlled store experiments (i.e., sales floors of Store A and Store B).

\paragraph{Article-To-Fixture Mapping Accuracy}
Similar to the controlled store environment, we also perform an evaluation of the article-to-fixture mapping accuracy.
However, in contrast to the previous setup, we do this with a much higher fixture granularity, due to the dynamic and unsupervised store environment (e.g., (parts of) fixtures can be rearranged and moved).
Specifically, articles are mapped to individual components of fixtures, such as the individual layers of a multilayered table.
Overall we use data from 6 stocktakes per store (i.e., 2 performed by store staff, and 4 additional stocktakes performed by us), collected in two consecutive weeks.
To calculate article-to-fixture mappings, we use our DBSCAN-based approach, as it showed a more robust performance in the controlled store experiments.
However, we use a smaller \( \epsilon \)-value to obtain more distinct clusters by DBSCAN.
Moreover, in this setup we do not leverage any historic information due to the limited number of performed stocktakes, and changes in article assortments between consecutive weeks.

We achieve an average accuracy of \( 69.0\% \) with a standard deviation of \( \sigma = 7.89\% \) over the 4 stocktakes that we perform in Store A.
On the other hand, the accuracy for stocktakes performed by store staff is only \( 21.57\% \) and \( 30.26\% \).
For stocktakes performed in Store B by us, we achieve an average accuracy of \( 85.48\% \) (\( \sigma = 1.64\% \)), while individual accuracies for stocktakes performed by staff are \( 53.17\% \) and \( 49.83\% \).

The lower accuracy compared to the controlled store setup (cf. Table~\ref{tab:results}) is a consequence of several factors. 
For example, the number of articles that have to be mapped is much higher in the pilot (i.e., on average \( 71\% \) more articles), due to the the fact that we also take article colors into account (i.e., potentially assigning the same article in a different color to a separate fixture), which makes the mapping more challenging.
Moreover, we find that the large gap between the stocktakes performed during the controlled experiments and the stocktakes performed by store staff in the pilot can most likely be attributed to different walking patterns.
We, for the most part subconsciously, walked along the fixtures in the store and scanned fixtures one-by-one, which better supports our approach when correlating reads from the articles on the fixtures with the reads of the corresponding reference tags.
Conversely, store staff additionally scans across the sales floor from a fixed position, where item tags and reference tags located all over the room respond, or scan multiple fixtures at once by constantly switching between them (i.e., scanning left and right while walking along an isle).
This introduces a lot of noise in the recorded data, which increases the challenge to properly map articles to fixtures with our methods.

\paragraph{Fixtures and Fixture Management}
The increased number of fixtures, and therefore on average smaller distances between fixtures, compared to the controlled store setup (i.e., on average \( 43.6\% \) more fixtures) also drastically increases the margin for error.
Nevertheless, we still find that the median distance to the correct fixture (i.e., error) is only \( 1 \) for wrongly assigned articles across the performed stocktakes.
Hence, we assign an article to a neighboring fixture in case of a wrong prediction in a large number of cases.

Moreover, due to the fact that many fixtures in both stores consist of multiple parts (e.g., multilayered tables), it is likely that we assign articles to the wrong part of a fixture, as reference tags belonging to different parts are in very close proximity.
To verify this, we perform an additional experiment, where we merge the individual parts of fixtures to one logical fixture (i.e., the two-layered table that originally constituted two fixtures is now aggregated to a single fixture). 
This reduces the number of fixtures on average by \( 17.4\% \) for Store A and by \( 36.8\% \) for Store B.
Consequently, we achieve an average accuracy of \(73.60\% \) (\(\sigma = 7.73\%\)) in Store A and \(93.41\%\) (\(\sigma = 1.95\%\)) in Store B over the stocktakes performed by us.
Further, accuracies for stocktakes performed by staff improve to \( 23.86 \% \) and \( 33.42\% \) in Store A, and to \( 61.62\% \) and \( 59.32\% \) in Store B.
This drastic increase in accuracy underlines the challenges related to this granular fixture management that is required in a real-world store setup. 

\paragraph{Article-To-Zone Mappings}
To minimize the impact of this issue, we experiment with article mappings on zones instead of fixtures.
Specifically, we divide Store~A in 8 and Store~B in 5 non-overlapping zones. 
These correspond to certain areas of a store (e.g., the clearance sale area), consisting, on average, of 6 individual fixtures.
We find that, according to the involved fashion retailer, a mapping based on zones is more common and more practical, as retailers often arrange their stores in thematic areas anyways (e.g., kid's fashion or footwear).

We achieve an average article-to-zone mapping accuracy of \( 89.22\% \) (\(\sigma = 4.43\%\)) in Store A and \( 96.63\% \) (\(\sigma = 1.47\%\)) in Store B.
Further, the accuracies of article mappings to zones based on store staff stocktakes are \( 53.55\% \) and \( 64.47 \% \) in Store A and \( 74.64\% \) and \( 90.84 \% \) in Store B.
While the increase in accuracies can likely be explained by an easier prediction task, reference tag management can be simplified significantly at the same time.
For example, reference tags can be placed on permanently installed fixtures or other appliances within a zone.
To illustrate this, we solely use tags placed on the walls of the sales floors to calculate article-to-zone mappings and achieve an average accuracy of \( 83.85 \% \) (\(\sigma = 3.91\%\)) in Store A and \( 84.14 \% \) (\(\sigma = 12.32\%\)) in Store B.
Hence, a detailed and extensive fixture management in a potentially highly dynamic store environment is not required anymore, while still maintaining a high zone mapping accuracy.
Note that the much lower accuracy in Store B can be attributed to an outlier stocktake, as indicated by the large standard deviation.

\paragraph{Reference Tags}
Independent of mapping articles to fixtures or zones in a brick-and-mortar store, the first and most critical step is the placement of reference tags.
This entails the availability of suitable RFID tags in stores as well as the process of encoding them.
While most stores maintain a contingent of empty (i.e., blank) tags that are typically used to replace malfunctioning or lost tags, we find that these do not suffice for our use-case.
For example, the fixtures in our real-world store setups are made of metal, which would negatively impact the readability of reference tags.

Therefore, we use ``on-metal'' tags made of hard plastic (see Figure~\ref{fig:on-metal-tag}) for the pilot that can be used on metallic surfaces without negatively affecting the readability of the tags.
The tags also have an adhesive strip and are already pre-encoded by the manufacturer, which makes their permanent placement on fixtures much easier.
While this allows for a more convenient and reliable placement of reference tags that is consistent across store setups (e.g., different fixture types and materials), it also increases the financial costs related to the introduction of an article-to-fixture mapping system.
Note that we verified that these ``on-metal'' tags are similar to the other RFID tags that we used in our experiments in terms of readability, to avoid adaptions to the presented methods.

\section{Conclusions \& Future Work}
In this paper we present a methodology to determine article-to-fixture assignments in fashion retail stores by leveraging read events collected during ordinary RFID-based stocktakes.

\paragraph{Proposed Methodology}
Specifically, we match reads of strategically placed reference tags with reads of items that are located in a store.
To that end, we propose two approaches to predict article-to-fixture assignments based on clustering of read events, and the similarity of time series generated based on the same data.
Moreover, we also leverage historic stocktake data to increase the accuracy of computed article-to-fixture assignments.
Using our methods we are not only able to assign articles to fixtures with an accuracy of more than 93\% in real-world store environments, but also limit the error in terms of distance to the correct fixture for wrong assignments.

\paragraph{Practical Implications}
Furthermore, when adapting this method in real-world retail stores we find that:

\begin{enumerate}[(i)]
    \item Special ``on-metal'' tags should be used as reference tags to cover heterogeneous environments and durability.
    \item A mapping from articles to zones simplifies fixture management drastically, while providing sufficient information about the location of articles for most retailers.
    \item During stocktakes store staff should, whenever possible, scan fixtures one-by-one to avoid fixture cross-reads.
\end{enumerate}

From our point of view, by taking these points into account only limited additional effort is required to gain information about the location of goods in real-world retail stores.
This builds the foundation for retailers to improve existing operational processes (e.g., replenishment or inventory reconciliation), or introduce new ones that generate valuable insights about their businesses and customers (e.g., Money Mapping).

\paragraph{Future Work}
So far, we only focused on the assignment of articles to fixtures, which requires the aggregation of read events of items belonging to the same article.
However, for future work we are also interested in applying our approach on individual items instead of articles.
This would open up additional use cases in stores, such as the detection of misplaced items on the sales floor.
Moreover, we are also interested to further investigate the best way of finding suitable parameter configurations for our approaches for individual stores, which do not depend on extensive parameter optimization.

The methodology we present in this paper highlights the potential of leveraging RFID-based stocktake data for additional use cases in modern retail stores.
Additionally, by publishing the data set we collected, we want to enable other researchers to develop methods for various applications in RFID-based retail environments.

\section*{Acknowledgments}
This work was supported by an industry-related dissertation scholarship granted by the Austrian Research Promotion Agency (FFG) under Grant Number 869522.

%Bibliography
\bibliographystyle{unsrt}  
\bibliography{references}

\end{document}